\documentclass{svmult}

\usepackage{makeidx}         % allows index generation
\usepackage{graphicx}        % standard LaTeX graphics tool
\usepackage{natbib}
\usepackage{multicol}        % used for the two-column index
\usepackage[bottom]{footmisc}% places footnotes at page bottom
\makeindex             % used for the subject index
\begin{document}

\title*{Cavity Microwave Searches for Cosmological Axions}

\author{Gianpaolo Carosi \inst{1} \and Karl van Bibber\inst{2}}
\institute{Lawrence Livermore National Laboratory,
 Livermore, CA, 94550, USA\\
\texttt{$^1$ carosi2@llnl.gov}\\
\texttt{$^2$ vanbibber1@llnl.gov}}

%\author{Gianpaolo Carosi \inst{1} \and Karl van Bibber\inst{2}}
%% Use \authorrunning{Short Title} for an abbreviated version of
%% your contribution title if the original one is too long
%\institute{Lawrence Livermore National Laboratory,
% Livermore, CA, 94550, USA
%\texttt{carosi2@llnl.gov}
%\and \texttt{vanbibber1@llnl.gov}}

\maketitle

\begin{abstract}
This chapter will cover the search for dark matter axions based on microwave cavity experiments 
proposed by Pierre Sikivie. The topic begins with a brief overview of halo dark matter and
the axion as a candidate. The principle of resonant conversion of axions in an external 
magnetic field will be described as well as practical considerations in optimizing the 
experiment as a signal-to-noise problem. A major focus of the lecture will be the two 
complementary strategies for ultra-low noise detection of the microwave photons - the
``photon-as-wave'' approach (i.e. conventional heterojunction amplifiers and soon to be
quantum-limited SQUID devices), and ``photon-as-particle'' (i.e. Rydberg-atom 
single-quantum detection). Experimental results will be presented; these experiments have
already reached well into the range of sensitivity to exclude plausible axion models, for 
limited ranges of mass. The section will conclude with a discussion of future plans and
challenges for the microwave cavity experiment.

%This section will cover the search for dark matter axions based on microwave cavity experiments 
%proposed by Pierre Sikivie. The topic begins with a brief overview of halo dark matter and
%the axion as a candidate. The principle of resonant conversion of axions in an external 
%magnetic field will be described as well as practical considerations in optimizing the 
%experiment as a signal-to-noise problem. The lecture will then focus on a technical description
%and results of the microwave cavity experiement ADMX currently running at Lawrence Livermore 
%National Laboratory, USA. Future experimental developements, such as lower noise amplifiers
%and Rydberg atom detection techniques will also be described.

\end{abstract}

\section{Dark matter and the axion}
\label{sec:1:new}

Recent precision measurements of various cosmological parameters have revealed a universe
in which only a small fraction can be observed directly. Measurements of deuterium abundances
predicted from the theory of big bang nucleosynthesis (BBN) have limited the 
familiar baryonic matter to a mere 4\% of the universe's total energy density \cite{BBN}. 
Evidence from the cosmic microwave background, combined with supernovae 
searches, galaxy surveys and other measurements leads to the fascinating conclusion that
the vast majority of the universe is made of gravitating ``dark matter'' (26\%) and a 
negative pressure ``dark energy'' (70\%) \cite{Tegmark}. 

Though the evidence for ``dark energy'' is relatively recent (primarily resting on 
cosmological supernovae surveys taken over the last decade or so) the existence of ``dark
matter'' has been known since the early 1930s. It was then that Fritz Zwicky, surveying the 
Coma cluster, noticed that member galaxies were moving far too quickly to be
gravitationally bound by the luminous matter \cite{Zwicky}. Either they were unbound, which meant the 
cluster should have ripped apart billions of years ago 
or there was a large amount of unseen ``dark matter'' keeping the system together. Since those
first observations evidence for 
dark matter has accumulated on scales as small as dwarf galaxies 
(kiloparsecs)\cite{JKG_DM_full} to the size of the observable universe 
(gigaparsecs)\cite{WMAP_1}. 

Currently the best dark matter candidates appear to be undiscovered non-baryonic 
particles left over from the big bang\footnote{Even without the limits from Big Bang 
Nucleosythesis searches for baryonic dark matter in cold gas clouds \cite{Gas} or MAssive Compact 
Halo Objects (MACHOs), like brown dwarfs \cite{Machos,EROS}, have not detected nearly enough to 
account for the majority of dark matter. Attempts to modify the laws of gravity at larger scales 
have also had difficulties matching observations \cite{MOND}.}. By 
definition they would have only the feeblest interactions with standard model particles 
such as baryons, leptons and photons. Studies of structure formation in the universe 
suggest that the majority of this dark matter is ``cold'', i.e. non-relativistic at the 
beginning of galactic formation. Since it is collisionless, relativistic dark matter would tend 
to stream out of initial density perturbations effectively smoothing out
the universe before galaxies had a chance to form \cite{Structure}. The galaxies that we 
observe today tend
to be embedded in large halos of dark matter which extend much further than their luminous
boundaries. Measurements of the Milky Way's rotation curves (along with other observables
such as microlensing surveys) constrain the density of dark matter near the solar system to 
be roughly $\rho_{CDM} \approx 0.45$ GeV/cm$^3$ \cite{Local_Density}.

The two most popular dark matter candidates are the general class of Weakly Interacting Massive 
Particles (WIMPs), one example being the supersymmetric neutralino, and the axion, 
predicted as a solution to the ``Strong CP'' problem. Though both particles are 
well motivated this discussion will focus exclusively on the axion. As described in earlier
chapters the axion is a light chargeless pseudo-scalar boson (negative parity, spin-zero particle) 
predicted from the breaking of the Peccei-Quinn symmetry. This symmetry was originally 
introduced in the late 1970s to explain why charge (C) and parity (P) appear to be conserved in 
strong interactions, even though the QCD Lagrangian has an explicitly CP violating term.
Experimentally this CP violating term should have lead to an easily detectable electric dipole 
moment in the neutron but none has been observed to very high precision \cite{Dipole}.

The key parameter defining most of the axion's characteristics is the spontaneous symmetry 
breaking (SSB) scale of the Peccei-Quinn symmetry, $f_a$. 
Both the axion couplings and mass are inversely proportional to $f_a$ with the mass defined as 

\begin{equation}
m_a \simeq 6\;\mu eV\left(\frac{10^{12}\;GeV}{f_a}\right).
\label{equ:Mass}
\end{equation}
and the coupling of axions to photons ($g_{a \gamma \gamma}$) expressed as

\begin{equation}
g_{a\gamma \gamma} \equiv \frac{\alpha g_{\gamma}}{\pi f_a}
\label{equ:Coupling}
\end{equation}
where $\alpha$ is the fine structure constant and $g_{\gamma}$ is a dimensionless model dependent 
coupling parameter. Generally $g_{\gamma}$ is thought to be $\sim 0.97$ for the class of
axions denoted KSVZ (for Kim-Shifman-Vainshtein-Zakharov) \cite{KSVZ_1,KSVZ_2} and $\sim -0.36$ for 
the more pessimistic grand-unification-theory inspired DFSZ (for 
Dine-Fischler-Srednicki-Zhitnitshii) models \cite{DFSZ_1,DFSZ_2}.
Since interactions are proportional to the square of the couplings these values of $g_{\gamma}$
tend to constrain the possible axion-to-photon conversion rates to only about an order of magnitude 
at any particular mass.

Initially $f_a$ was believed to be around the electroweak scale ($f_a \sim 250\;GeV$) resulting 
in an axion mass of order $100\;keV$ \cite{Weinberg,Wilczek} and couplings strong enough to be seen 
in accelerators. Searches for axions in particle and nuclear experiments, along with 
limits from astrophysics, soon lowered its possible mass to 
$m_a \leq 3 \times 10^{-3}\;eV$\cite{Cavity_review}
corresponding to $f_a \geq 10^9\;GeV$. Since their couplings are inversely proportional to 
$f_a$ these low mass axions were initially thought to be undetectable and were 
termed ``invisible'' axions.

From cosmology it was found that a general lower limit could be placed on the axion mass as
well. At the time of the big bang 
axions would be produced in copious amounts via various mechanisms described in previous 
chapters. 
The total contributions to the energy density 
of the universe from axions created via the vacuum misalignment method can then be expressed as

\begin{equation}
\Omega_a \sim \left(\frac{5\;\mu eV}{m_a}\right)^{7/6}
\label{Overclosure}
\end{equation}
which puts a lower limit on the axion mass of $m_a \geq 10^{-6}\;eV$ (any lighter and the
axions would overclose the universe, $\Omega_a \geq 1$). 
Combined with the astrophysical and experimental limits this results in a 3 decade mass range 
for the axion, from $\mu eV\;-\;meV$, with the lower masses more likely if the axion is the 
major component of dark matter. The axions generated in the early universe around the 
QCD phase transition, when the axion mass turns on, would have momenta $\sim 10^{-8}\;eV/c$ 
while the surrounding plasma had a temperature $T \simeq 1\;GeV$ \cite{Cavity_review}. 
Furthermore, such axions are so weakly coupling that they would never be in thermal equilibrium 
with anything else.
This means they would constitute non-relativistic ``cold'' dark matter 
from the moment they appeared and could start to form structures around density perturbations 
relatively quickly. 

Today the axion dark matter in the galaxy would consist of a large halo of particles moving with 
relative velocities of order $10^{-3}c$. It is unclear whether any or all of the axions would be
gravitationally thermalized but, in order for them to be bound in the galaxy, they would have to be 
moving less than the local escape velocity of $2\times 10^{-3}c$.
It's possible that
non-thermalized axions could still be oscillating into and out of the galaxy's gravitational well.
These axions would have extraordinarily tiny velocity dispersions (of order $10^{-17}c$) 
\cite{Caustics} and the
differences in velocity from various infalls (first time falling into the galaxy, first time flying 
out, second time falling in, etc.) would be correlated with the galaxy's development.

\section{Principles of microwave cavity experiments}
\label{sec:2:new}

Pierre Sikivie was the first to suggest that the ``invisible'' axion could actually be 
detected \cite{Cavity_idea}. This possibility rests on the coupling of axions to photons given
by

\begin{equation}
L_{a \gamma \gamma} = -\left(\frac{\alpha}{\pi} \frac{g_{\gamma}}{f_a}\right)a \vec{E} \cdot \vec{B}
\label{Lagrangian}
\end{equation}
where $\vec{E}$ and $\vec{B}$ are the standard electric and magnetic field of the coupling photons, 
$\alpha$ is the
fine structure constant and $g_{\gamma}$ is the model dependent coefficient mentioned in the 
previous section \cite{Cavity_review}. 
Translating this to a practical experiment Sikivie suggested that axions
passing through an electromagnetic cavity permeated with a magnetic field could resonantly 
convert into photons when the cavity resonant frequency ($\omega$) matched the axion mass 
($m_a$). Since the entire mass of the axion would be converted into a photon a 5 $\mu eV$ 
axion at rest would convert to a 1.2 GHz photon which could be detected with sensitive 
microwave receivers. The predicted halo axion velocities of order $\beta = 10^{-3}$ would 
predict a spread in the axion energy, from $E_a = m_a c^2 + \frac{1}{2}m_a c^2\beta^2$, of 
order $10^{-6}$. For our example 5 $\mu eV$ axions this would translate into a 1.2 kHz 
upward spread in the frequency of converted photons. The power of
axions converting to photons on resonance in a microwave cavity is given by

\begin{eqnarray}
P_a & = & \left(\frac{\alpha}{\pi} \frac{g_{\gamma}}{f_a}\right)^2 V B^2_0 \rho_a C_{lmn} \frac{1}{m_a} Min\left(Q_L,Q_a\right) \label{equ:Power}\\
    & = & 0.5\times 10^{-26}W \left(\frac{V}{500\;liters}\right)\left(\frac{B_0}{7\;T}\right)^2 C \left(\frac{g_{\gamma}}{0.36}\right)^2 \nonumber\\ 
& & \times \left(\frac{\rho_a}{0.5\times 10^{-24} g/cm^3}\right) \nonumber\\ 
& & \times \left(\frac{m_a}{2\pi\;(GHz)}\right)Min\left(Q_L,Q_a\right) \nonumber
\end{eqnarray}
where $V$ is the cavity volume, $B_0$ is the magnetic field, $Q_L$ is the cavity's loaded 
quality factor (defined as center frequency over frequency bandwidth), $Q_a=10^6$ is the 
quality factor of the axion signal (axion energy over spread in energy or $1/\beta^2$), 
$\rho_a$ is the axion mass density at the detection point (earth) and $C_{lmn}$ is the form factor 
for 
one of the transverse magnetic ($TM_{lmn}$) cavity modes (see section \ref{subsec:3b:new} for more 
on cavity modes). This form factor is essentially the
normalized overlap integral of the external static magnetic field, $\vec{B_0}(\vec{x})$, and 
the oscillating electric field, $\vec{E}_{\omega}(\vec{x})e^{i\omega t}$, of that particular cavity 
mode. It can be determined using 

\begin{equation}
C = \frac{|\int_V d^3x \vec{E}_{\omega}\cdot \vec{B}_0|^2}{\vec{B}^2_0 V \int_V d^3x \epsilon |\vec{E}_{\omega}|^2}
\label{equ:Form_factor}
\end{equation}
where $\epsilon$ is the dielectric constant in the cavity.

For a cylindrical cavity with a homogeneous longitudinal magnetic
field the $TM_{010}$ mode provides the largest form factor ($C_{010}=0.69$)
\cite{Cavity_review}. Though model dependent equation 
\ref{equ:Power} can give an idea of the incredibly small signal, measured in yoctowatts 
($10^{-24}$ W), expected from axion-photon conversions in a resonant cavity.
This is much smaller than 
the $2.5\times 10^{-21}\;W$ of power received from the last signal of the $Pioneer\;10$ 
spacecraft's 7.5-W transmitter in 2002, when it was 12.1 billion kilometers from earth 
\cite{Pioneer}.

Currently the axion mass is constrained between a $\mu eV$ and a $meV$ corresponding to 
a frequency range for converted photons between 240 MHz and 240 GHz. To maintain the resonant
quality of the cavity, however, only a few kHz of bandwidth can be observed at any one time. As a
result the cavity needs to be tunable over a large range of frequencies in order to cover all 
possible values of the axion mass. This is accomplished using metallic or dielectric tuning rods 
running the length of the cavity cylinder. Moving the tuning rods from the edge to the center of 
the cavity shifts the resonant frequency by up to 100 MHz.

Even when the cavity is exactly tuned to the axion mass detection is only possible if the 
microwave receiver is sensitive enough to distinguish the axion conversion signal over the 
background noise from the cavity and the electronics. 
The signal to noise ratio (SNR) can be calculated from the Dicke radiometer equation 
\cite{Dicke_radiometer}

\begin{equation}
SNR = \frac{P_a}{\bar{P}_N} \sqrt{Bt} = \frac{P_a}{k_B T_S}\sqrt{\frac{t}{B}}
\label{Radiometer}
\end{equation}
where $P_a$ is the axion conversion power, $\bar{P}_N=k_B B T_S$ is the average thermal noise 
power, $B$ is the bandwidth, $T_S$ is the total system noise temperature (cavity plus electronics) 
and $t$ is the signal integration time 
\cite{Cavity_review}. With the bandwidth of the experiment essentially set
by the axion mass and anticipated velocity dispersion ($\beta^2 \sim 10^{-6}$) the SNR can be
raised by either increasing the signal power ($P_a \propto B^2_0 V$), lowering the noise 
temperature or integrating for a longer period of time. Increasing the size of the magnetic field
or the volume of the cavity to boost the signal power can get prohibitively expensive fairly 
quickly. Given the large range of possible masses the integration time needs to remain relatively
short (of order 100 seconds integration for every kHz) in order to scan an appreciable amount in 
time scales of a year or so. If one chooses a specific SNR that would be acceptable for detection 
then a scanning rate can be defined as

\begin{eqnarray}
\frac{df}{dt} & = & \frac{12\;GHz}{yr} \left(\frac{4}{SNR}\right)^2\left(\frac{V}{500\;liter}\right)\left(\frac{B_0}{7\;T}\right)^4 \label{equ:Rate}\\
 & & \times\;C^2\left(\frac{g_{\gamma}}{0.36}\right)^4\left(\frac{\rho_a}{5\times10^{-25}}\right)^2 \nonumber\\
 & & \times \left(\frac{3K}{T_S}\right)^2\left(\frac{f}{GHz}\right)^2\frac{Q_L}{Q_a}. \nonumber
\end{eqnarray}

Given that all other parameters are more or less fixed, due to physics and budgetary constraints, the
sensitivity of the experiment (both in coupling reach and in scanning speed) can only practically be
improved by developing ultra low noise microwave receivers. In fact some of the quietest microwave
receivers in the world have been developed to detect axions \cite{Med_res}.

\section{Technical implementation}
\label{sec:3:new}

The first generation of microwave experiments were carried out at Brookhaven National Laboratory
(BNL) \cite{Brookhaven} and at the University of Florida \cite{UofF} in the mid-1980s. These
were proof-of-concept experiments and got within factors of 100 - 1000 of the sensitivity 
required to detect plausible dark matter axions (mostly due to their small cavity size and 
relatively high noise temperatures) \cite{Cavity_review}. In the early 1990s second generation
cavity experiments were developed at Lawrence Livermore National Laboratory (LLNL) in the U.S.
and in Kyoto, Japan. Though both used a microwave cavity to convert the axions to photons
they each employed radically different detection techniques. The U.S. experiment 
focused on improving coherent microwave amplifiers (photons as waves) while the Japan experiment
worked to develop a Rydberg-atom single-quantum detector (photons as particles). Since the Kyoto
experiment is still in the development phase we will save its description for a later section and
focus on the U.S. experiment. 

\begin{figure}
\centering
\includegraphics[height=8.5cm]{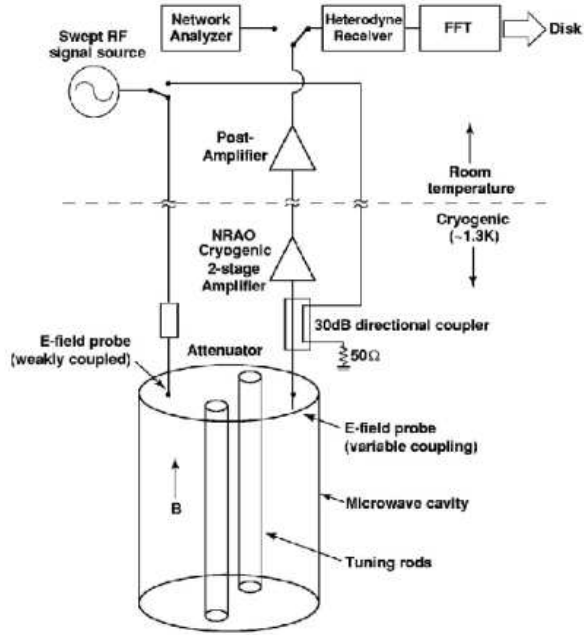}
%\picplace{5cm}{2cm} % Give the correct figure height and width in cm
%
\caption{Schematic diagram of ADMX experiment including both the resonant cavity (which sits in
the bore of a superconducting solenoid) and receiver electronics chain.}
\label{fig:ADMX_schem}       % Give a unique label
\end{figure}
A schematic of the LLNL experiment, dubbed the \textbf{A}xion \textbf{D}ark \textbf{M}atter
e\textbf{X}periment (ADMX), can be seen in figure \ref{fig:ADMX_schem}.
The experiment consists of a cylindrical copper-plated steel cavity containing two axial tuning rods.
These can be moved transversely from the edge of the cavity wall to its center allowing one to 
perturb the resonant frequency. The cavity itself is located in the bore of a 
superconducting solenoid providing a 
strong constant axial magnetic field. The electromagnetic field of the cavity is coupled to low-noise
receiver electronics via a small adjustable antenna\cite{Cavity_review}. These electronics
initially amplify the signal using two ultra-low noise cryogenic amplifiers arranged in series. 
The signal is then boosted again via a room temperature post-amplifier
and injected into a double-heterodyne receiver.
The receiver consists of an image reject mixer to reduce the signal frequency from 
the cavity resonance (hundreds of MHz - GHz) to an intermediate frequency (IF) of 10.7 MHz. 
A crystal bandpass filter is then employed to reject noise power outside of a 35 kHz window
centered at the IF. Finally the signal is mixed down to almost audio frequencies (35 kHz) and 
analyzed by fast-Fourier-transform (FFT) electronics which compute a 
50 kHz bandwidth centered at 35 kHz. Data is taken every 1 kHz or so by moving the tuning rods 
to obtain a new resonant $TM_{010}$ mode. In the next few sections we will expand on 
some of these components.

\begin{figure}
\centering
\includegraphics[height=8.5cm]{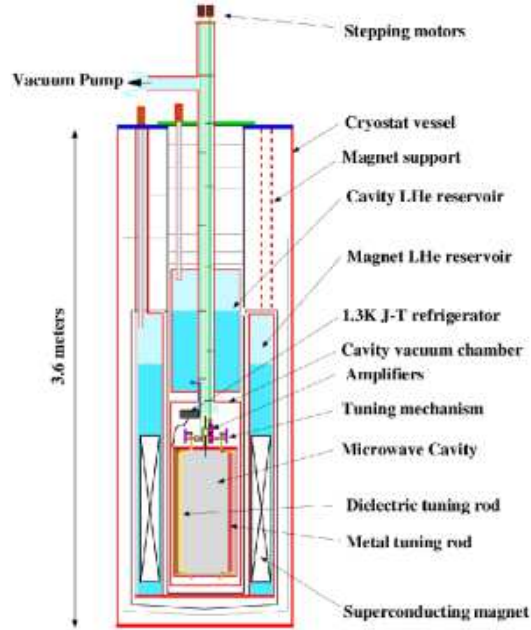}
%\picplace{5cm}{2cm} % Give the correct figure height and width in cm
%
\caption{Overview of ADMX hardware including superconducting magnet and cavity insert.}
\label{fig:ADMX_exp}       % Give a unique label
\end{figure}

\subsection{Magnet}
\label{subsec:3a:new}

The main magnet for ADMX was designed to maximize the $B^2_0 V$ contribution to the signal power
(equation \ref{equ:Power}). It was determined that a superconducting solenoid would yield the 
most cost effective solution and its extremely large inductance (535 Henry) would have the added 
benefit of keeping the field 
very stable. The 6 ton magnet coil is housed in a 3.6 meter tall cryostat 
(see figure \ref{fig:ADMX_exp}) with an open magnet bore allowing the experimental insert,
with the cavity and its liquid helium (LHe) reservoir, to be lowered in. 
The magnet itself is immersed during operations in a 4.2 K LHe bath in order 
to keep the niobium-titanium windings superconducting. Generally the magnet was kept at a field 
strength of 7.6 T in the solenoid center (falling to approximately 70 \% strength at the ends) 
but recently its been run as high as 8.2 T \cite{Cavity_review}. 

\subsection{Microwave cavities}
\label{subsec:3b:new}

The ADMX experiment uses cylindrical cavities in order to maximize the axion conversion 
volume in the solenoid bore. They are made of a copper-plated steel cylinder
with capped ends. The electromagnetic field structure inside a cavity can be found by solving the
Helmholtz equation 

\begin{equation}
\nabla^2\Phi + k^2\Phi = 0
\label{equ:Helmholtz}
\end{equation}
where the wavenumber $k$ is given by 

\begin{equation}
k^2 = \mu \epsilon \omega^2 - \beta^2
\label{equ:wavenumber}
\end{equation}
and $\beta$ is the eigenvalue for the transverse (x,y) component \cite{Kinion_thesis}. The cavity
modes are the standing wave solutions to equation \ref{equ:Helmholtz}. The boundary conditions of an 
empty cavity only
allow transverse magnetic (TM) modes ($\vec{B}_z = 0$) and transverse electric (TE) modes
($\vec{E}_z = 0$). Since the TE modes have no axial electric field one can see from equation
\ref{Lagrangian} that they don't couple at all to axions and we'll ignore them for the moment. 
The $TM_{lmn}$ modes are three dimensional standing waves 
where $l=0,1,2...$ is the number of azimuthal nodes, $m=1,2,3...$ is the number of radial
nodes and $n=0,1,2...$ is the number of axial nodes. The axions couple most strongly to the lowest
order $TM_{010}$ mode.

The resonant frequency of the $TM_{010}$ mode can be shifted by the introduction of metallic
or dielectric tuning rods inserted axially into the cavity. Metallic rods
raise the cavity resonant frequency the closer they get to the center while dielectric rods
lower it. In ADMX these rods are attached to the ends of alumina arms which pivot about axles set in
the upper and lower end plates. The axles are rotated via stepper motors mounted at the top of the
experiment (see figure \ref{fig:ADMX_exp}) which swing the tuning rods from the cavity edge to 
the center in a circular arc. The stepper motors are attached to a gear reduction which translates
a single step into a 0.15 arcsecond rotation, corresponding to a shift of $\sim$ 1 kHz at 
800 MHz resonant frequency \cite{Cavity_review}. 

\begin{figure}
\centering
\includegraphics[height=6.cm]{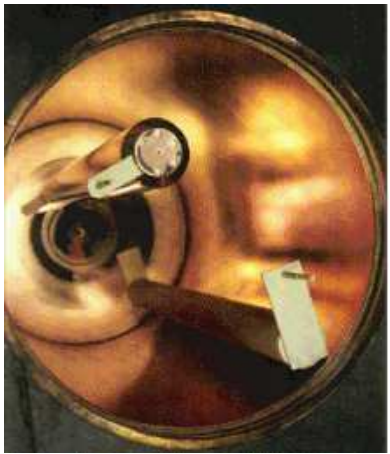}
%\picplace{5cm}{2cm} % Give the correct figure height and width in cm
%
\caption{Resonant cavity with top flange removed. An alumina tuning rod can be seen at the bottom
right and a copper tuning rod is in the upper left.}
\label{fig:Cavity}       % Give a unique label
\end{figure}

With the addition of metallic tuning rods TEM modes ($\vec{B}_z = \vec{E}_z = 0$) can also be 
supported in the cavity. Like the TE modes they do not couple to the axions but they can couple
weakly to the vertically mounted receiver antenna (due to imperfections in geometry, etc). 
\begin{figure}
\centering
\includegraphics[height=5.cm]{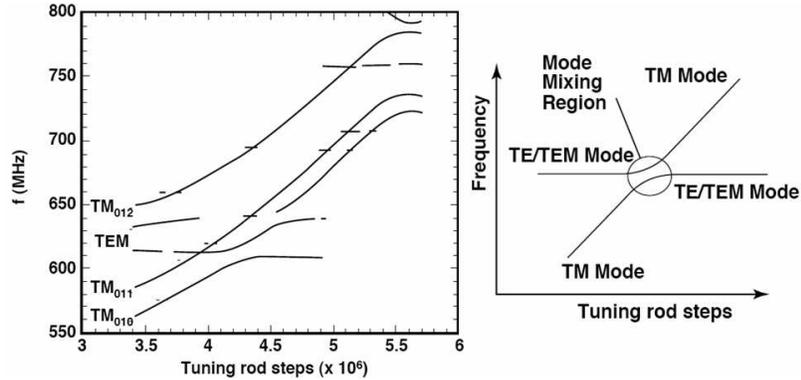}
%\picplace{5cm}{2cm} % Give the correct figure height and width in cm
%
\caption{Mode structure of a cavity with two copper tuning rods. The left figure displays
the frequencies of the resonant modes, 
measured via a swept rf signal, when one tuning rod is kept at the cavity edge while the other is
moved toward the center. The right figure is a sketch of a mode crossing.}
\label{fig:Cavity_Modes}       % Give a unique label
\end{figure}
Figure \ref{fig:Cavity_Modes} demonstrates how the various resonant modes
shift as a copper tuning rod is moved from near the cavity wall toward the center. The TEM and
TE modes are largely unaffected
by the change in tuning rod position while TM modes
rise in frequency as one of the copper rods moves toward the cavity center. This leads to regions
in which a TM mode crosses a TE or TEM mode (referred to as mode mixing). These  mode mixings 
(illustrated by the 
right part of figure \ref{fig:Cavity_Modes}) introduce frequency gaps which can not be scanned. As
a result the cavity was later filled with LHe, which changed the microwave 
index of refraction to 1.027, thus lowering the mode crossings by 2.7\% and allowed the previously
unaccessible frequencies to be scanned.

A key feature of the resonant microwave cavity is its quality factor $Q$, which is a measure of 
the sharpness of the cavity response to external excitations. It is a dimensionless value 
which can be defined a number of ways including the ratio of the stored energy ($U$) to the 
power loss ($P_L$) per cycle: $Q = \omega_0 U/P_L$. 
The quality factor (Q) of the $TM_{010}$ mode is determined by sweeping a radio (rf) signal through
the weakly coupled antenna in the cavity top plate (see figure \ref{fig:ADMX_schem}). Generally, 
the unloaded Q of the cavity is $\sim 2\times 10^5$ \cite{Cavity_review} which is very near to 
the theoretical maximum for oxygen-free annealed copper at cryogenic temperatures. During data 
taking the insertion depth of
the major antenna is adjusted to make sure that it matches the 50 $\Omega$ impedance of the 
cavity (called critically coupling). When the antenna is critically coupled half the microwave
power in the cavity enters the electronics via the antenna while half is dissipated in the cavity
walls. Overcoupling the cavity would lower the Q and thus limit the signal enhancement while 
undercoupling the cavity would limit the microwave power entering the electronics. 

\subsection{Amplifier and receiver}
\label{subsec:3c:new}

After the axion signal has been generated in the cavity and coupled to the major port antenna it is
sent to the cryogenic amplifiers. The design of the first amplifier is especially important 
because its noise temperature (along with the cavity's Johnson noise) dominates the rest of 
the system. 
This can be illustrated by following a signal from the cavity as 
it travels through two amplifiers in series. The power contribution from the thermal noise of the 
cavity at temperature $T_c$ over bandwidth $B$ is given by $P_{nc} = B k_B T_c$ (where $k_B$ is
Boltzmann's constant). When this 
noise passes through the first amplifier, which provides gain $G_1$, the output includes the 
boosted cavity noise as well as extra power ($P_{N,A_1}$) from the amplifier itself. The noise 
from the amplifier appears as an increase in the temperature of the input source.
\begin{equation}
P_1 = G_1 B k_B T_c + P_{N,A_1} = G_1 B k_B (T_c + T_{A_1})
\label{equ:first_stage}
\end{equation}
If this boosted noise power (cavity plus first amplifier) is then sent through a second amplifier,
with gain $G_2$ and noise temperature $T_{A_2}$, the power output becomes
\begin{equation}
P_2 = G_2 P_1 + P_{N,A_2} = G_2 (G_1 B k_B (T_c + T_{A_1})) + G_2 B k_B T_{A_2}
\label{equ:noise_amps}
\end{equation}
The combined noise temperature from the two amplifiers ($T_A$) can be found by matching equation
\ref{equ:noise_amps} to that of a single amplifier, $P_2 = G_2 G_1 B k_B (T_c + T_A)$, 
which gives
\begin{equation}
T_A = T_{A_1} + \frac{T_{A_2}}{G_1}
\label{equ:amp_temp}
\end{equation}
Thus one can see that, because of the gain $G_1$ of the first stage amplifier, its noise temperature
dominates all other amplifiers in the series. 

\begin{figure}
\centering
\includegraphics[height=6.cm]{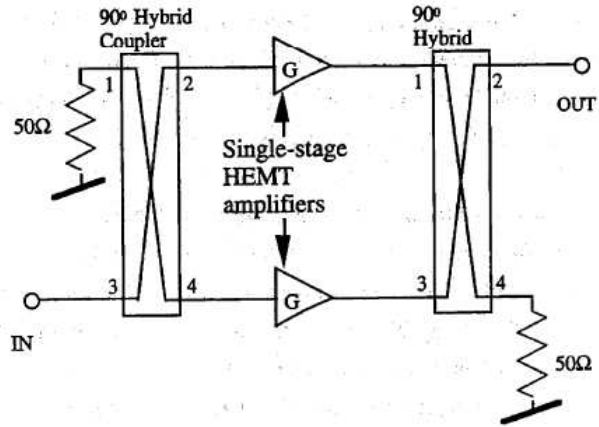}
%\picplace{5cm}{2cm} % Give the correct figure height and width in cm
%
\caption{Schematic diagram of a balanced amplifier. Every time the signal crosses through the 
middle of a hybrid its phase is shifted by 90 degrees. Reflections back to the input destructively
interfere while reflections to the upper left constructively interfere and are dumped into a 50 
$\Omega$ terminator. Signals to the output are both shifted by 90 degrees and thus add 
constructively.}
\label{fig:Balanced}       % Give a unique label
\end{figure}

The current first stage amplifiers used in ADMX are cryogenic heterostructure field-effect 
transistors (HFETs) developed at the National Radio Astronomy Observatory (NRAO) specifically for 
the ADMX experiment \cite{Cavity_review,NRAO}. In these amplifiers 
electrons from an aluminum doped gallium arsenide layer fall into the GaAs two-dimensional 
quantum well (the FET channel). The FET electrons travel ballistically, with little scattering,
thus minimizing electronic noise \cite{Physics_Today}. Currently electronic noise temperatures of
under 2 K have been achieved using the HFETs. In the initial ADMX data 
runs, now concluded, two HFET amplifiers were used in series, each with approximately 17 dB power 
gain, leading to a total first stage power gain of 34 dB. Each amplifier
utilized 90 degree hybrids in a balanced configuration in order to minimize input reflections,
thus providing a broadband match to the 50 $\Omega$ cavity impedance (see figure 
\ref{fig:Balanced}).

Though the amplifiers worked well in the high magnetic field just above the cavity it was 
determined during commissioning that they should be oriented such that the magnetic field 
was parallel to the HFET channel electron flow. This minimized the electron travel path and thus
the noise temperature \cite{Cavity_review}. 

The signal from the cryogenic amplifiers is carried by coaxial cable to a low-noise room 
temperature post-amplifier, which added an additional 38 dB gain between 300 MHz - 1 GHz. Though
the post-amplifiers noise temperature is 90 K its contribution relative to the cryogenic 
amplifiers (with 38 dB initial gain) is only 0.03 K (see equation \ref{equ:amp_temp}). 
Including various losses the total gain from the cavity to the post-amplifier output is 69 dB
\cite{Cavity_review}.

\begin{figure}
\centering
\includegraphics[height=4.cm]{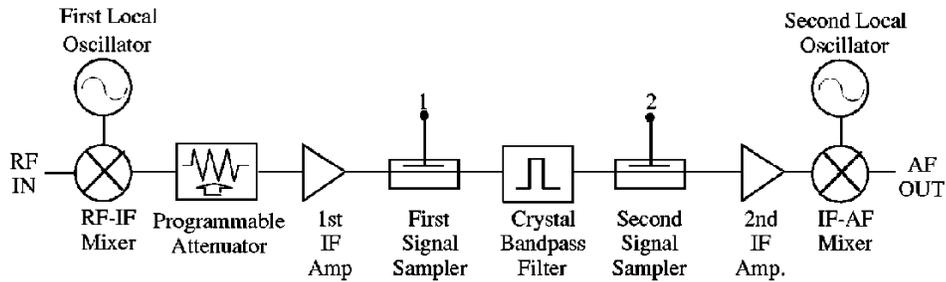}
%\picplace{5cm}{2cm} % Give the correct figure height and width in cm
%
\caption{Receiver chain that mixes the signal down from the cavity $TM_{010}$ resonant frequency 
to 35 kHz.}
\label{fig:Receiver}       % Give a unique label
\end{figure}

After initial stages of amplification the signal enters the double-heterodyne receiver (essentially
an AM radio). Figure \ref{fig:Receiver} is a schematic of the receiver electronics. The first 
element is an image reject mixer which uses a local oscillator 
to mix the signal down to 10.7 MHz. This intermediate frequency (IF) is then sent through a 
programmable attenuator (used during room temperature testing so that the 
receiver electronics are not saturated). An IF amplifier then boosts the 
signal by another 20 dB before passing it by a weakly coupled signal sampler. The signal then 
passes through a crystal bandpass filter which suppresses noise outside a 30 kHz bandwidth center 
at 10.7 MHz. The signal is then boosted by an additional 20 dB before being mixed down to 35 kHz. 
The total amplification of the signals from the cavity is $\sim$ 106 dB \cite{General_art_1}.

Once the signal has been mixed down to the 35 kHz center frequency it is passed off to a commercial 
FFT spectrum analyzer and the power spectrum recorded. The entire receiver, including filter, is
calibrated using a white-noise source at the input. 
During data collection the FFT
spectrum analyzer takes 8 msec single-sided spectra (the negative and positive frequency components
are folded on top of each other). Each spectrum consists of 400 bins with 125 Hz width spanning a 
frequency range of 10-60 kHz. After 80 seconds of data taking (with a fixed cavity mode) the 
10,000 spectra are averaged together and saved as raw data. This is known as the medium resolution
data.

In addition there is a high resolution channel to search for extremely narrow conversion
lines from late infall non-thermal axions (as mentioned at the end of section \ref{sec:1:new}).
For this channel the 35 kHz signal is passed through a passive LC filter with a 6.5 kHz passband, 
amplified, 
and then mixed down to a 5 kHz center frequency. A single 
spectrum is then obtained by acquiring $2^{20}$ points in about 53 sec and a FFT is performed.
This results in about $3.4\times 10^5$ points in the 6.5 kHz passband with a frequency resolution of 
19 mHz. 

\section{Data analysis}
\label{sec:4:new}

The ADMX data analysis is split into medium and high resolution channels. The medium resolution
channel is analyzed using two hypotheses. The first is a ``single-bin'' search motivated by the 
possibility that some of the axions have not thermalized and therefore would have negligible 
velocity dispersion, thus depositing all their power into a single power-spectrum bin. The
second hypothesis utilizes a ``six-bin'' search which assumes that axions have a velocity dispersion 
of order $10^{-3}c$ or less (axions with velocities greater than $2\times 10^{-3}c$ would escape the 
halo). The six-bin search is the most conservative and is valid regardless of whether the halo 
axions have thermalized or not. 

Since each 80 second long medium resolution spectra is only shifted by 1 kHz from the previous
integration each frequency will show up in multiple spectra (given the 50 kHz window). As a result
each 125 Hz bin is weighted according to where it falls in the cavity response function
and co-added to give an effective integration time of $\sim$ 25 minutes 
per frequency bin. For the single-bin search individual 125 Hz bins are selected if they 
exceed an initial power-level threshold. This is set relatively low so a large number of bins are 
usually selected. These bins are then rescanned to achieve a similar signal-to-noise ratio
and combined with the first set of data generating a spectra with higher signal to noise. The 
selection process is then repeated a number of times until persistent candidates are identified.
These few survivors are then carefully checked to see if there are any external sources of 
interference that could mimic an axion signal. If all candidates turn out to be exterior radio 
interference the excluded
axion couplings (assuming a specific dark matter density) can be computed from the near-Gaussian 
statistics of the single-bin data. For the six-bin search, all six adjacent frequency bins that
exceed a set power-threshold are selected from the power spectra. The large number of candidates
are then whittled down using the same iterations as the single-bin analysis. If no candidates survive
the excluded axion couplings are computed by Monte Carlo \cite{Cavity_review}.

From the radiometer equation (\ref{Radiometer}) one can see that the search sensitivity can be 
increased if strong narrow spectral lines exist. The integration times for each tuning rod setting 
is around 60 seconds and the resulting Doppler shift from the Earth's 
rotation leads to a spread of $\sim$ mHz in a narrow axion signal. Since the actual velocity 
dispersions of each discrete flow is unknown multiple resolution searches were performed by 
combining 19 mHz wide bins. These were referred to as $n$-bin searches, where $n = $ 1, 2, 4, 8, 64, 
512 and 4096. Candidate peaks were kept if they were higher then a specified
threshold set for that particular $n$-bin search. These thresholds were 20, 25, 30, 40, 120, 650 and
4500 $\sigma$, for increasing order of $n$. The initial search using the high resolution
analysis took data between 478-525 MHz, corresponding to axion masses between 1.98 and 2.17 $\mu eV$.
This search was made in three steps. First the entire frequency range was scanned in 1 kHz 
increments with the candidate axion peaks recorded. Next multiple time traces were taken of 
candidate peaks \cite{High_res}. Finally persistent peaks were checked by attenuating or disconnecting various 
diagnostic coaxial cables leading into the cavity (see figure \ref{fig:ADMX_schem}). If the 
signals were external interference they would decrease in power dramatically while
an axion signal would remain unchanged \cite{Cavity_review}. Further checks could be done by 
disconnecting the cavity from the receiver input and replacing it with an antenna to see if
the signal persisted.

If a persistent candidate peak is found which does not have an apparent
source from external interference a simple check would be to turn off the magnetic field. If the 
signal disappears it would be a strong indication that it was due to axions and not some unknown
interference. So far, though, all candidates have been identified with an external source.

\section{Results}
\label{sec:5:new}

\begin{figure}
\centering
\includegraphics[height=5.5cm]{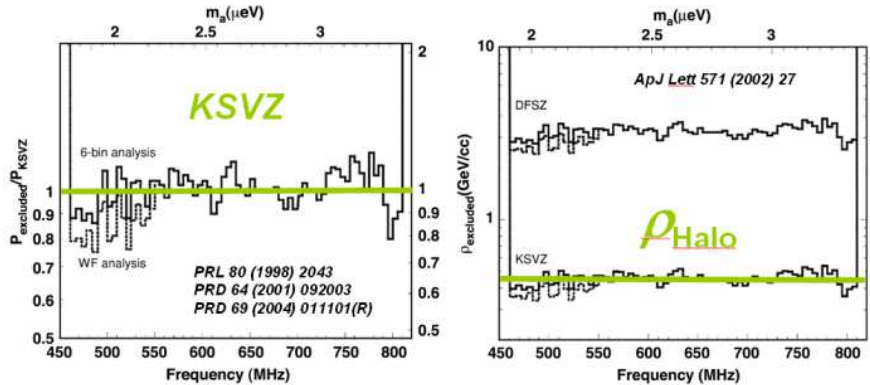}
%\picplace{5cm}{2cm} % Give the correct figure height and width in cm
%
\caption{Results from the medium resolution channel \cite{Med_res}. The figure to the left is the 
exclusion plot for power in a thermalized spectrum assuming a halo density of $\rho_a = 0.45\;GeV/cm^3$. 
The figure to the right is the fractional dark matter halo density excluded as axions for two different 
axion models.}
\label{fig:Med_res_limits}       % Give a unique label
\end{figure}

\begin{figure}
\centering
\includegraphics[height=5.cm]{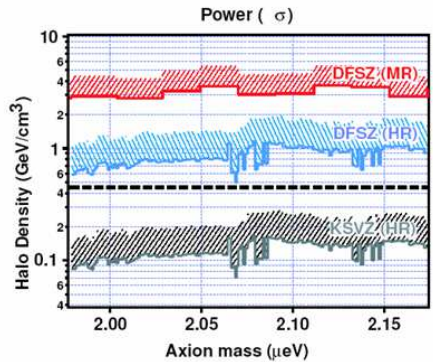}
%\picplace{5cm}{2cm} % Give the correct figure height and width in cm
%
\caption{High resolution limits given different axion couplings \cite{High_res}. This shows that 
the current high resolution channel is sensitive to fractional halo densities ($\approx$ 30\%) if 
the axions couple via the KSVZ model. If they couple via the DFSZ model the experiment is not yet 
sensitive to the maximum likelihood halo density ($\rho_a \sim 0.45\;GeV/cm^3$), but would be 
sensitive to a single line with twice that density.}
\label{fig:High_res_limits}       % Give a unique label
\end{figure}

So far no axions have been detected in any experiment. ADMX currently provides the 
best limits from microwave cavity experiments in the lowest mass range (most plausible if axions 
are the major component to the dark matter). Both the medium resolution data and the high 
resolution data yield exclusion plots in either the coupling strength of the axion (assuming a 
halo density of $\rho_a = 0.45\;GeV/cm^3$) or in the axion halo density (assuming a specific 
DFSZ or KSVZ coupling strength). Results from the medium resolution channel \cite{Med_res} 
can be seen in figure \ref{fig:Med_res_limits} and the high resolution results \cite{High_res} 
can be seen in \ref{fig:High_res_limits}. Both of the results are listed at 90 \% confidence level. 

\section{Future developments}
\label{sec:6:new}

In order to carry out a definitive search for axion dark matter various improvements to the detector
technology need to be carried out. Not only do the experiments need to become sensitive enough to 
detect even the most pessimistic axion couplings (DFSZ) at fractional halo densities but they must
be able to scan relatively quickly over a few decades in mass up to possibly hundreds of GHz. 
The sensitivity of the detectors (which is also related to scanning speed) is currently limited by
the noise in the cryogenic HFET amplifiers. Even though they have a noise temperature under 2 K the 
quantum limit (defined as $T_Q \sim h\nu/k$) is almost two orders of magnitude lower (25 mK at 
500 MHz). To get down to, or even past, this quantum limit two very different technologies are being
developed. The first is the implementation of SQUIDs (Superconducting Quantum Interference Devices) 
as first stage cryogenic amplifiers. The second uses Rydberg-atoms
to detect single microwave photons from axion conversions in the cavity. 

Though both techniques 
will lead to vastly more sensitive experiments they will still be limited in their mass range. 
Currently all cavity experiments have been limited to the $2 - 20 \mu eV$ range, mostly due to the
size of resonant cavities. For a definitive search the mass range must be increased by a factor of 
50 which requires new cavity designs that increase the resonant frequency while maintaining large 
enough detection volumes. Detectors that work at these higher frequencies also need to be developed.

\subsection{SQUID amplifiers}
\label{sec:6a:new}

\begin{figure}
\centering
\includegraphics[height=3.5cm]{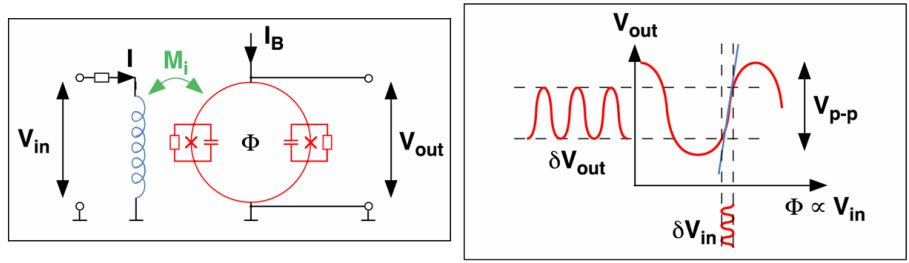}
%\picplace{5cm}{2cm} % Give the correct figure height and width in cm
%
\caption{Essentials of a SQUID microwave detector. The left figure is a schematic of the SQUID 
device coupling to the input signal which is converted into magnetic flux. The right figure shows
how biasing the flux allows for amplification.}
\label{fig:SQUID}       % Give a unique label
\end{figure}

The next generation of the ADMX experiment will use SQUID amplifiers to replace the first stage
HFETs. SQUIDs essentially use a superconducting loop with two parallel Josephson
junctions to enclose a total amount of magnetic flux $\Phi$. This includes both a fixed flux 
supplied by the bias coil and the signal flux supplied by an input coil. The phase difference
between the currents on the two sides of the loop are affected by changing $\Phi$ resulting in an
interference effect similar to the two-slit experiment in optics \cite{Kinion_thesis}.  Essentially 
the SQUID will act as flux to voltage transducers as illustrated in figure \ref{fig:SQUID}.

\begin{figure}
\centering
\includegraphics[height=3.5cm]{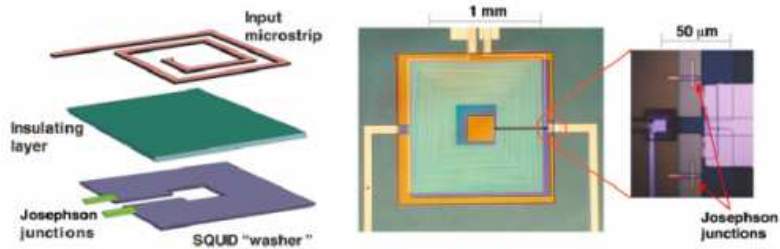}
%\picplace{5cm}{2cm} % Give the correct figure height and width in cm
%
\caption{Diagram and picture of microstrip resonator SQUIDs to be used in ADMX upgrade.}
\label{fig:SQUID_res}       % Give a unique label
\end{figure}

Most SQUIDs are built using the Ketchen and Jaycox design \cite{Ketchen}, in which the SQUID loop
is an open square washer made of niobium (Nb). The loop is closed by a separate Nb electrode 
connected to the washer opening on either side by a Josephson junction and external shunt resistors.
A spiral input coil is placed on top of the washer, separated by a layer of insulation. The 
original designs in which input signals were coupled into both ends of the coil tended to only 
work below about 200 MHz due to parasitic capacitance between the coil and the washer at higher 
frequencies. This was solved by coupling the input signal between one end of the coil and the 
SQUID washer, which would act as a ground plane to the coil and create a microstrip resonator 
(see figure \ref{fig:SQUID_res}). This design has been tested successfully up to 3 GHz
\cite{Kinion_thesis}.

Unlike the HFETs, whose noise temperature bottoms out at just under 2 K regardless of how cold the
amplifiers get, the SQUIDs noise temperature remains proportional to the physical temperature down
to within 50\% of the quantum limit. The source of this thermal noise comes from the shunt resistors 
across the SQUID's Josephson junctions and future designs that minimize this could push the noise 
temperature even closer to the quantum limit \cite{Physics_Today}. 

Currently the ADMX experiment is in the middle of an upgrade in which SQUIDs will be installed as
first stage cryogenic amplifiers. This should cut the combined noise temperature of the cavity + 
electronics in half allowing ADMX to become sensitive to half the KSVZ coupling (with the same 
scanning speed as before). Due to the SQUIDs' sensitivity to magnetic fields this upgrade 
includes an entire redesign in which a second superconducting magnet is being installed in order to 
negate the main magnet's field around the SQUID amplifiers. Data taking
is expected to begin in the first half of 2007 and run for about a year.
Future implementations of ADMX foresee using these SQUID detectors with a dilution refrigerator
to set an operating temperature of $\sim$ 100 mK, allowing sensitivity to DFSZ axion couplings to 
be achieved with 5 times the scanning rate the current HFETs take to reach KSVZ couplings.

\subsection{Rydberg-atom single-quantum detectors}
\label{sec:6b:new}

One technique to evade the quantum noise limit is to use Rydberg atoms to 
detect single photons from the cavity. A Rydberg atom has a single valence electron promoted to a
level with a large principal quantum number $n$. These atoms have energy spectra similar in many
respects to hydrogen, and dipole transitions can be chosen anywhere in the microwave spectrum by
an appropriate choice of $n$. The transition energy itself can be finely tuned by using the Stark
effect to exactly match a desired frequency. That, combined with the Rydberg atom's long lifetime 
and large dipole transition probability, make it an excellent microwave photon detector.

\begin{figure}
\centering
\includegraphics[height=6.cm]{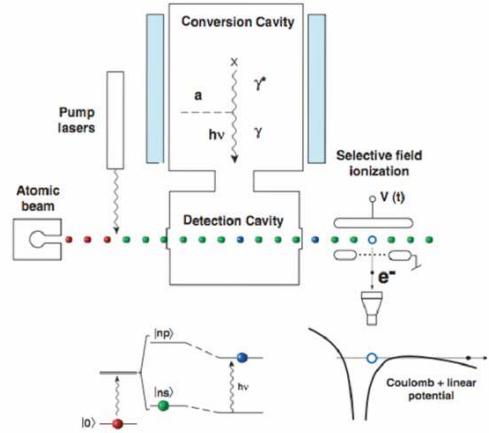}
%\picplace{5cm}{2cm} % Give the correct figure height and width in cm
%
\caption{Schematic of single photon microwave detection utilizing Rydberg atoms.}
\label{fig:Rydberg-atom}       % Give a unique label
\end{figure}

An experimental setup utilizing this technique called CARRACK has been assembled in Kyoto, Japan 
\cite{Cavity_review,Kyoto_1} and a schematic can be seen pictured in figure \ref{fig:Rydberg-atom}. 
The axion
conversion cavity is coupled to a second ``detection'' cavity tuned to the same resonant frequency
$\nu$. A laser excites an atomic beam (in this case rubidium) into a Rydberg state 
($|0> \rightarrow |n>$) which then 
traverses the detection cavity. The spacing between the energy levels is adjusted to $h\nu$
using the Stark effect and microwave photons from the cavity can be efficiently absorbed by the 
atoms (one photon per atom, $|ns> \rightarrow |np>$). The atomic beam then exits the cavity and is 
subjected to selective field ionization in which electrons from atoms in the higher energy state 
($|np>$) get just enough energy to be stripped off and detected \cite{Physics_Today}.

Currently the Kyoto experiment has measured cavity emission at 2527 MHz down to a temperature
of 67 mK, a factor of two below the quantum limit at that frequency, and is working to reach the
eventual design goal of 10 mk \cite{Kyoto_1}. This would be the point in which the cavity blackbody 
radiation 
would become the dominant noise background. One deficiency of the Rydberg atom technique is that
it can't detect structure narrower than the bandpass ($\Delta E/E$) of the cavity (generally 
$\sim 10^{-5}$). As a result it is insensitive to axion halo models that predict structure down
to $\Delta E/E \sim 10^{-11}$, an area in which the ADMX high resolution channel, utilizing 
microwave amplifiers, can cover. Despite this Rydberg atom detectors could become very useful tools
for halo axion detection in the near future.

\subsection{Challenge of higher frequencies}
\label{sec:6c:new}

Current microwave cavity technology has only been able to probe the lowest axion mass scale. In 
order to cover the entire range up to the exclusion limits set by SN 1987a of 
$m_a \leq meV$ new cavity and detection techniques must be investigated which can operate up to 
the 100 GHz range. The resonant cavity frequency essentially depends on the size the cavity and the
resonant mode used. The $TM_{010}$ mode has by far the largest form factor ($C\sim 0.69$) of any mode
and all other higher frequency modes have much smaller or identically zero form factors. The single
50 cm diameter cavity used in the initial ADMX experiments had a central resonant frequency 
($TM_{010}$) of 460 MHz and radial translation of metallic or dielectric tuning rods could only 
raise or lower that frequency by about $\pm 50\%$ \cite{Cavity_review}. Smaller cavities could 
get higher frequencies but the rate of axion conversions would go down as the cavity volume
decreased. 

\begin{figure}
\centering
\includegraphics[height=6.cm]{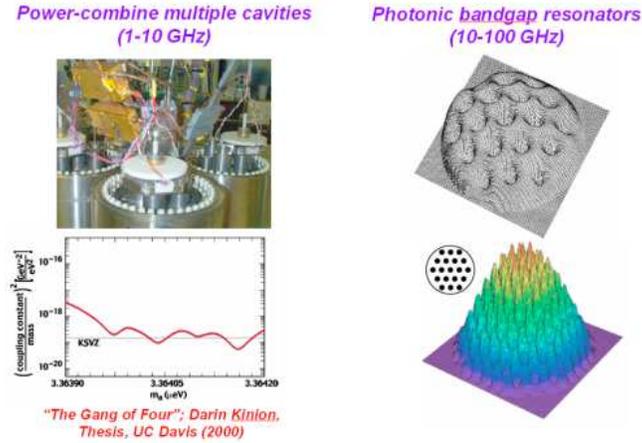}
%\picplace{5cm}{2cm} % Give the correct figure height and width in cm
%
\caption{Outline of possible cavity concepts to explore higher axion masses. The left figure 
includes both a picture of the 4 cavity array and its corresponding exclusion plot over 
the limited mass range it took data. The right figure includes field maps for multiple posts 
inserted in a cavity.}
\label{fig:Higher_frequency}       % Give a unique label
\end{figure}

In order to use the full volume of the magnet with smaller cavities it was determined that 
multiple cavities could be stacked next to each other and power combined. 
As long as the de Broglie wavelength of the axions is larger than the total array individual 
cavities tuned to the same frequency can be summed in phase.
Typical axion 
de Broglie wavelengths are $\lambda_{dB} \sim 10m - 100m$ which means they drive the $\sim 1m$ 
cavity volume coherently. Data taken using a four cavity array in ADMX reached KSVZ sensitivity 
\cite{Kinion_thesis} over 
a small mass range (see figure \ref{fig:Higher_frequency}). These initial tests 
had difficulties getting the piezoelectric motors working trouble free in the magnetic and 
cryogenic environment. 
Since those tests the technology has advanced to the point in which it may be feasible to create 
larger sets of smaller cavity areas.

To reach even higher frequencies ideas have been raised to use resonators with periodic arrays of
metal posts. Figure \ref{fig:Higher_frequency} shows the electric field profile of one possible
array using a 19 post hexagonal pattern. Mounting alternating posts from the cavity top and the 
bottom
and translating them relative to each other allow the resonant frequency to be adjusted by 10\%
or so. The possibility of using such cavities, or other new cavity geometries, is an active area
of research and progress needs to be made before the full axion mass range can be explored.

\section{Summary, conclusions}
\label{sec:7:new}

Experimentally the axion is a very attractive cold dark matter candidate. Its coupling to 
photons ($g_{\gamma}$) for several different models all fall within about an order of magnitude 
in strength and its mass scale is currently confined to a three decade window. This leaves the axion
in a 
relatively small parameter space, the first two decades or so of which is within reach of current 
or near future technology.

The ADMX experiment has already begun to exclude dark matter axions with KSVZ couplings over the lowest
masses and upgrades to SQUID amplifiers and a dilution refrigerator could make ADMX sensitive
to DFSZ axion couplings 
over the first decade in mass within the next three years. Development of advanced Rydberg-atom detectors,
along with higher frequency cavities geometries, could 
give rise to the possibility of a definitive axion search within a decade. By 
definitive we mean a search which would either detect axions at even the most pessimistic couplings
(DFSZ) at fractional halo densities over the full mass range, or rule them out entirely. 

It should be noted that if the axion is detected it would not only solve the Strong-CP problem 
and perhaps the nature of dark 
matter but could offer a new window into astrophysics, cosmology and quantum physics. Details of 
the axion spectrum, especially if fine structure is found, could provide new information of how the 
Milky Way was formed. The large size of the axions de Broglie wavelength 
($\lambda_a \sim 10m - 100m$) could even allow for interesting quantum experiments to be 
performed at macroscopic scales. All of these tantalizing possibilities, within the 
reach of current and near future technologies, makes the axion an extremely exciting dark matter
candidate to search for.

\section{Acknowledgments}

This work was supported under the auspices of the U.S. Department of Energy
under Contract W-7405-Eng-48 at Lawrence Livermore National Laboratory.

%%%%%%%%%%%%%%%%%%%%%%%%%%%%%%%%%%%%%%%%%%%%%%%%%%%%%%%%%%%%%%%%%%%%%
% BibTeX users please use
 \bibliographystyle{unsrt}
 \bibliography{carosi}
%
% Non-BibTeX users please follow the syntax
% the syntax of "referenc.tex" for your own citations
%\input{referenc}
%%%%%%%%%%%%%%%%%%%%%%%%%%%%%%%%%%%%%%%%%%%%%%%%%%%%%%%%%%%%%%%%%%%%%%

%%%%%%%%%%%%%%%%%%%%%%%%%%%%%%%%%%%%%%%%%%%%%%%%%%%%%%%%%%%%%%%%%%%%%%

\printindex
\end{document}